\documentclass[10pt,journal,compsoc]{IEEEtran}

\usepackage{graphicx}

\ifCLASSOPTIONcompsoc
  \usepackage[nocompress]{cite}
\else
  \usepackage{cite}
\fi

\begin{document}

\title{The Discovery of Mutated Driver Pathways in Cancer: Models and Algorithms}

\author{Junhua~Zhang and~Shihua~Zhang 
\IEEEcompsocitemizethanks{\IEEEcompsocthanksitem J. Zhang and S. Zhang are with National Center for Mathematics and Interdisciplinary Sciences, Academy of Mathematics and Systems Science, Chinese Academy of Sciences, Beijing 100190, China. \protect\\
E-mail: zjh@amt.ac.cn, zsh@amss.ac.cn.}
}

\markboth{}
{Zhang \MakeLowercase{\textit{et al.}}: The Discovery of Mutated Driver Pathways in Cancer: Models and Algorithms}

%


\IEEEtitleabstractindextext{%
\begin{abstract}
The pathogenesis of cancer in human is still poorly understood. With the rapid development of high-throughput sequencing technologies, huge volumes of cancer genomics data have been generated. Deciphering those data poses great opportunities and challenges to computational biologists. One of such key challenges is to distinguish driver mutations, genes as well as pathways from passenger ones. Mutual exclusivity of gene mutations (each patient has no more than one mutation in the gene set) has been observed in various cancer types and thus has been used as an important property of a driver gene set or pathway. In this article, we aim to review the recent development of computational models and algorithms for discovering driver pathways or modules in cancer with the focus on mutual exclusivity-based ones.
\end{abstract}

\begin{IEEEkeywords}
Bioinformatics, cancer genomics, driver gene, driver pathway, mutual exclusivity, co-occurring mutation
\end{IEEEkeywords}}

\maketitle

\IEEEdisplaynontitleabstractindextext

%
\IEEEpeerreviewmaketitle

\section{Introduction}

\IEEEPARstart{U}{nderstanding} the mechanism of carcinogenesis has been a great challenge for human. With the rapid advance in deep sequencing technologies, several large-scale cancer projects have generated an unprecedented amount of cancer genomics data (e.g., The Cancer Genome Atlas (TCGA) ~\cite{TCGA_08,TCGA_11}, the International Cancer Genome Consortium (ICGC) ~\cite{ICGC}, the Cancer Cell Line Encyclopedia (CCLE) ~\cite{Barretina}, and the Therapeutically Applicable Research to Generate Effective Treatments (TARGET) ~\cite{Mullighan}). The rapid accumulation of huge volumes of genomic data has provided tremendous opportunities for the better understanding on cancer initiation, progression and development. Deciphering those data poses great challenges and computational problems for the community of bioinformatics and computational biology ~\cite{Jiang_14,Hofree,Hansen,Jiang_15}. For example, modeling cancer evolution is one of such key problems. Cancer is viewed as a somatic evolutionary process characterized by the accumulation of mutations ~\cite{Yates}. Evolutionary modeling for cancer describes the dynamics of tumor cell populations and makes inference about the evolutionary history of a tumor from molecular level ~\cite{Sun,Jiang_14,Wang_J}. Tumor stratification is another fundamental issue. The key is to classify heterogeneous tumor population into clinically and biologically meaningful subtypes by similarity of molecular profiles ~\cite{Lu_05,Reis-Filho,Hofree,Liu}. It has been demonstrated that tumor stratification has a profound implication for the individualization of efficient treatments of cancer patients ~\cite{Lu_05}.

Similarly, distinguishing driver mutations which contribute to cancer development, from passenger mutations that have accumulated in somatic cells but without functional consequences is a key challenge in computational cancer genomics. Distinguishing drivers from passengers in cancer can help to identify carcinogenic mechanisms and drug targets. In this review, we first briefly survey the methods for identifying driver genes, and we then review the models and algorithms of discovering driver pathways or modules in detail with the focus on mutual exclusivity-based ones using mutation data as well as others.

\section{Brief Survey of Methods for Identifying Driver Genes}

A common type of methods for identifying driver genes is based on gene mutational frequency (termed as frequency-based methods) ~\cite{Ding_08}. The basic principle of such methods is to test individual genes whether they are mutated in a significant number of cancer patients than expected by chance. A key step is to estimate the background mutation rate (BMR) to quantify the accumulation of random passenger mutations. Proper estimation of BMR is a key factor affecting the power of this type of methods. An overestimation of BMR fails to identify true recurrent mutations (false negatives), whereas an underestimation would lead to too many false positives. Early frequency-based methods assume a single constant background rate across the genome for all samples ~\cite{Ding_08}. However, recent studies demonstrat that BMR is not constant across the genome ~\cite{Youn_11,Lawrence}. Moreover, a number of features (other than mutation frequency) could affect the mutation rate including mutation types, sequence context ~\cite{Sjoblom}, gene-specific features ~\cite{Stamatoyannopoulos,Chen_2010}, mutation-specific scores that assess functional impact ~\cite{Ng_PC,Adzhubei,Reva} and so on.

Therefore, recent studies have developed a number of frequency-based methods which adopt one or more of these features to get a more accurate BMR estimation. For example, both MuSiC ~\cite{Dees} and MutSigCV ~\cite{Lawrence} employ the mutation types and sample-specific mutation rates. MutSigCV also allows for the inclusion of gene-specific features such as the expression level and replication timing. Youn and Simon ~\cite{Youn_11} considered mutation types and functional impacts of mutations in their approach. Recently, Vogelstein \emph{et al.} ~\cite{Vogelstein_13} investigated the spatial patterns of mutations within driver genes based on an integrative analysis across multiple cancers in COSMIC. They demonstrated that many known oncogenes and tumor suppressors possess non-random mutational patterns. Several methods adopt these new features with improved sensitivity and specificity ~\cite{Youn_11,Tamborero,Korthauer}. OncodriveCLUST uses the evidence of positional clustering to identify oncogenes ~\cite{Tamborero}. MADGiC (Model-based Approach for identifying Driver Genes in Cancer) is an unified empirical Bayesian model-based approach which uses all the above features to identify driver genes ~\cite{Korthauer}.

As we know, many mutations occur in different genes among different patients. Such mutational heterogeneity in cancer genomes is another important factor affecting the performance of frequency-based methods. This heterogeneity may be a consequence of the presence of passenger mutations in each cancer genome. It may also be a consequence of the cancer evolution due to the diverse mutational events during this process ~\cite{Nik-Zainal}.

Although individual tumors exhibit diverse genomic alterations, many studies have demonstrated that driver mutations tend to affect a limited number of cellular signaling and regulatory pathways ~\cite{Vogelstein_04,TCGA_08,Ding}. Thus, a common alternative to single gene test is to evaluate the recurrence of mutations in groups of genes derived from known pathways or genome-scale gene interaction networks. These groups of genes may be candidate driver pathways, which may be frequently perturbed within tumor cells and can lead to the acquisition of carcinogenic properties such as cell proliferation, angiogenesis, or metastasis ~\cite{Hanahan_00,Hanahan_11}.

Gene Set Enrichment Analysis (GSEA) can be employed to rank the list of mutated genes first, and then assess whether a pre-defined set of genes (such as a given pathway) has more high-ranking genes than would be expected by chance ~\cite{Subramanian}. Depending on different scoring techniques to rank genes, different approaches coupled with GSEA have been used to determine enrichment of mutations in certain pathways or cellular functions (such as CaMP-GSEA ~\cite{Lin_07}). Furthermore, considering the complex heterogeneity of cancer genomes, several GSEA-based methods have been presented by scoring each gene set at the patient level rather than the gene level ~\cite{Boca,Wendl}. These patient-oriented methods are more interpretable and statistically powerful than traditional gene-oriented methods.

It has long been realized that driver mutations or genes perturb signaling, regulatory or metabolic pathways that promote the development and progression of cancer. Thus, many biological pathway or interaction network based methods have been developed to identify significantly mutated subnetworks. We have known that the mutational landscape of cancer consists of `mountains' of a few frequently mutated genes and `hills' of many more less frequently mutated genes ~\cite{Vogelstein_13,Garraway}. One advantage of the network-based methods is that it can detect less frequently mutated driver genes which are members of gene sets recurrently mutated ~\cite{Wu_10,Vandin_11,Leiserson_15}. Vandin \emph{et al.} developed a network-based method (HotNet) ~\cite{Vandin_11} and employed a heat diffusion process on the interaction network to define a local neighborhood of `influence' for each mutated gene, and then identified some recurrently mutated subnetworks with a two-stage multiple hypothesis test. More recently, they further extended it (HotNet2) ~\cite{Leiserson_15} to identify pathways and protein complexes perturbed by somatic aberrations across multiple types of cancers. NetBox ~\cite{Cerami_10} and MEMo ~\cite{Ciriello} are two alternative methods. NetBox was developed based on the detection of closely connected network modules or communities. Notably, MEMo uses the strategy of mutual exclusivity of gene mutations to detect mutated subnetworks critical to carcinogenesis ~\cite{Ciriello}.

\begin{figure*}[!t]
\centering
\includegraphics[width=4.8in]{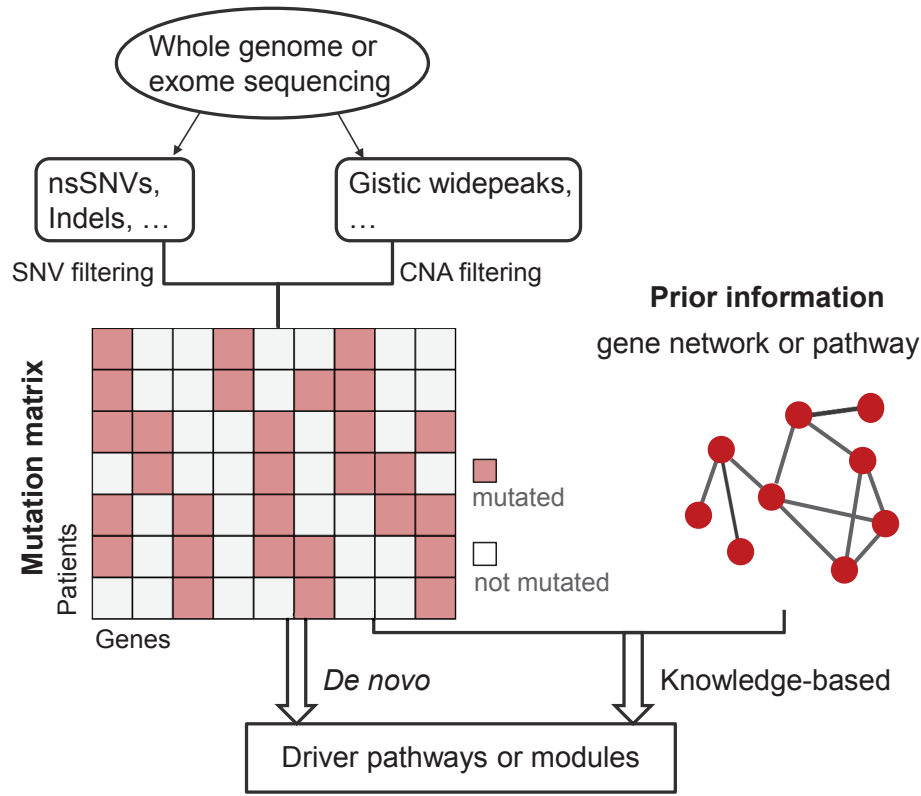}
\caption{An overview of the methods to discover driver pathways or core modules based on mutual exclusivity. Given a mutation matrix obtained from the DNA sequencing data, the methods of discovering driver pathways or core modules are classified into two types depending on whether they use the prior knowledge or not.}
\label{fig1}
\end{figure*}

The mutual exclusivity phenomenon that each patient has no more than one mutation in a gene set was observed in various cancer types ~\cite{Yeang}. For example, the mutation of \emph{TP53} and the copy number amplification of \emph{MDM2} seldom appear simultaneously in glioblastoma multiforme (GBM) patients (\emph{p53} pathway) ~\cite{TCGA_08}, and \emph{BRCA1/2} mutations and \emph{BRCA1} epigenetic silencing in serous ovarian cancer possess the similar property ~\cite{TCGA_11}. It has been commonly thought that mutually exclusive genomic events provide strong genetic evidence that the altered genes are functionally linked in a common biological pathway. Therefore, many approaches based on mutual exclusivity have been developed to identify driver pathways or core modules in recent years (Fig.~\ref{fig1}).

\section{Methods for Identifying Driver Pathways}

In this section, we review the methods for identifying driver pathways from three aspects: individual driver pathway, cooperative driver pathways and driver pathways across multiple types of cancers (pan-cancer level) (Table~\ref{table1}). Prior knowledge-based and \emph{de novo} identification methods are two major types for identifying driver pathways. We first introduce the simple mutual exclusivity-based pairwise search for mutational patterns (PSMP) ~\cite{Yeang}, which provides the basis for understanding many other methods.

\begin{table*}[!ht]
\renewcommand{\arraystretch}{1.3}
\caption{Brief Summary of Methods for Discovering Driver Pathways or Core Modules}
\label{table1}
\centering
    \begin{tabular}{l l l l}
      \hline
      Category & Method & Website & Reference \\
      \hline
      Individual driver pathway & & & \\
       $\;$ --Prior knowledge & & & \\
       & PSMP  & NA$^1$                         & ~\cite{Yeang}    \\
       & MEMo  & http://cbio.mskcc.org/memo     & ~\cite{Ciriello} \\
       & Mutex & http://code.google.com/p/mutex & ~\cite{Babur}    \\
       $\;$ --\emph{De novo} identification & & & \\
       & Dendrix   & http://compbio.cs.brown.edu/projects/dendrix/     & ~\cite{Vandin} \\
       & MDPFinder & http://page.amss.ac.cn/shihua.zhang/software.html & ~\cite{Zhao} \\
       & ME        & NA   & ~\cite{Szczurek} \\
       & RME       & http://brl.bcm.tmc.edu/rme/index.rhtml & ~\cite{Miller} \\
       & iMCMC     & NA   & ~\cite{Zhang_2013} \\
       & SODP$^2$  & http://pitttransmed-tcga.dbmi.pitt.edu/mutuallyExclusive/ & ~\cite{Lu2015} \\
       Cooperative pathways & & & \\
       & CoMDP     & http://page.amss.ac.cn/shihua.zhang/software.html & ~\cite{Zhang_2014} \\
       & Multi-Dendrix & http://compbio.cs.brown.edu/projects/multi-dendrix/ & ~\cite{Leiserson_13} \\
       & GAMToC$^3$ & http://sourceforge.net/p/melamedgamtoc & ~\cite{Melamed} \\
       & LM$^4$     & NA & ~\cite{Remy_2015} \\
      Pan-cancer analysis & & & \\
       & HotNet2$^4$ & http://compbio.cs.brown.edu/projects/hotnet2/ & ~\cite{Leiserson_15} \\
       & MEMCover$^4$ & NA & ~\cite{Kim_15} \\
      \hline
    \end{tabular}
\begin{flushleft}
    Several methods fall into multiple categories but are listed only once for simplicity.\\
    $^1$ NA: not applicable.\\
    $^2$ SODP uses GO annotations for differential expression genes.\\
    $^3$ GAMToC actually detects one gene set each time, in which some genes are mutually exclusive and some others are co-occurring.\\
    $^4$ LM, HotNet2 and MEMCover employ gene interaction networks.
\end{flushleft}
\end{table*}

\subsection{PSMP}

Yeang \emph{et al.} ~\cite{Yeang} examined the patterns of somatic mutations of cancers obtained from COSMIC in 45 different tissue types. They categorized the mutational patterns of multiple genes into two types in terms of co-occurrence and mutual exclusivity. They adopted the likelihood ratio (LR) statistic between the empirical frequency of co-occurrence and its expected frequency to test the combinatorial patterns of gene mutations. That is, for two genes $g_1$ and $g_2$, the statistic was defined as $\frac{P(g_1^m,g_2^m)} {P(g_1^m)P (g_2^m)}$, where $g_1^m$ ($g_2^m$) indicates the gene $g_1$ ($g_2$) is mutated and a low score suggests mutual exclusivity and a high score indicates co-occurrence of two given genes, respectively.

They obtained 105 significant combinatorial mutational patterns, which cover genes in six major pathways relevant to cancer including cell cycle control, stress response, \emph{Ras}, insulin growth factor (\emph{IGF-AKT}), \emph{Wnt}, and \emph{TGF}-$\beta$ signaling pathways. Most co-occurring patterns contain genes in different pathways, whereas most mutually exclusive pairs are in the same pathways. This result is consistent with the idea that cancer progression can be viewed as a Darwinian evolutionary process ~\cite{Vogelstein_88, Cahill}. Mutations of two genes participating in the same pathway or biological process rarely confer a significant selective advantage compared to the single mutation. By contrast, if the genes participate in different pathways or functions, their mutations may provide an additive or even synergistic role in conferring an advantage to the tumor. Therefore, one would expect to observe a tendency of mutually exclusive mutations of genes in the same pathway and the tendency of co-occurring mutations of genes participating in different pathways. This suggests mutual exclusivity can be a basic criterion or constraint for identifying driver pathways in cancer.

\subsection{Discovery of Individual Driver Pathway}

\subsubsection{Prior Knowledge-based Methods}

\begin{figure*}[!t]
\centering
\includegraphics[width=4.8in]{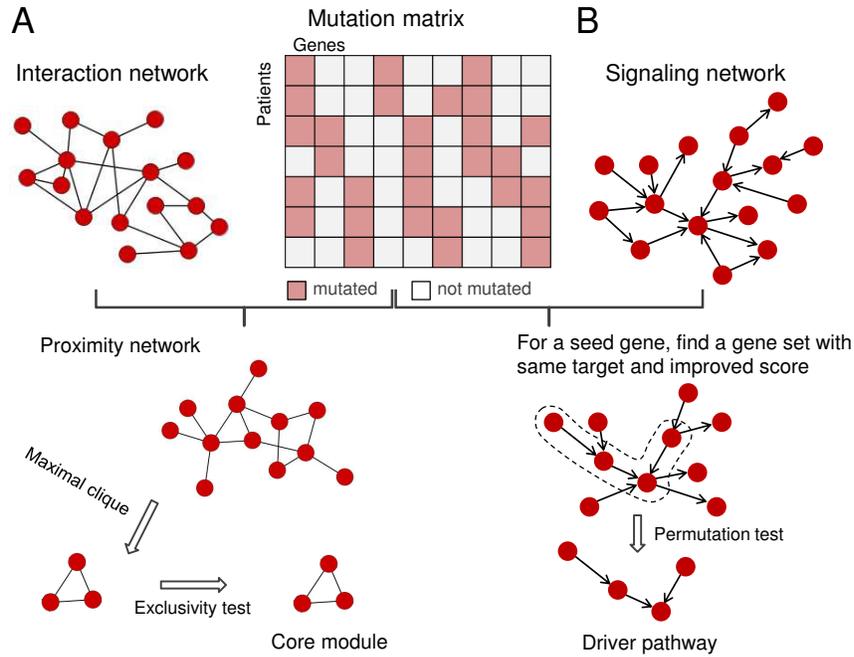}
\caption{A schematic illustration of the prior knowledge-based methods for identifying driver pathways. (A)	The strategy adopted in MEMo ~\cite{Ciriello}: assess the common neighbors of two genes and build a proximity network by combining an undirected interaction network and the mutation matrix; then extract maximal cliques and test the significance of mutual exclusivity of the mutations in the gene sets using a switching permutation principle. (B) The strategy adopted in Mutex ~\cite{Babur}: combine a prior directed signaling network and the mutation matrix; for any altered gene as a seed, apply a greedy algorithm to sequentially add candidate genes which have a common downstream target, and obtain a driver pathway with a FDR score.}
\label{fig2}
\end{figure*}

\noindent As stated in the previous section, some prior gene interaction networks can be used for the task of identifying driver pathways. The goal is to find a subnetwork in which the genes have statistically significant mutually exclusive mutations (Fig.~\ref{fig2}).

\vskip3mm
\noindent\textbf{MEMo}

\noindent Ciriello \emph{et al.} designed MEMo ~\cite{Ciriello} to identify candidate driver subnetworks with three properties: (1) the member genes of a driver pathway are recurrently altered across multiple patients; (2) the member genes tend to participate in the same pathway or biological process, and (3) alteration events within a driver pathway are mutually exclusive. Specifically, this method consists of four steps. First, for all the somatic mutations and copy number alterations (CNAs) across the observed samples, three gene filters are used to identify the statistically significantly mutated genes and significant regions of interest with concordant mRNA expressions which is represented as a binary event matrix. Second, MEMo performs a global gene comparison to assess the network proximity of two genes by assessing the number of their common neighbors based on prior pathway and network knowledge. Notably, although two genes do not connect each other, they can also be assessed as proximal if they share a large number of common neighbors. Third, a graph based on the pair proximity is built with an edge between two genes if their proximity is high. MEMo extracts all maximal cliques from this graph, which represent local clusters and may contain proteins of likely similar biological functions. And fourth, each detected clique is tested whether its member genes are mutually exclusive by a Markov chain Monte Carlo permutation strategy. MEMo has been effectively used in applications with moderate numbers of mutated genes, but a limitation is that it is not able to run on very large data (such as the pan-cancer data set) ~\cite{Leiserson_15,Babur}.

\vskip3mm
\noindent\textbf{Mutex}

\noindent Babur \emph{et al.} introduced a novel statistical metric to quantify the mutual exclusivity between more than two altered genes ~\cite{Babur}. By combining prior pathway knowledge with this statistic, they developed a new approach (Mutex) to identify groups of mutually exclusively altered genes that have a common downstream target. Specifically, they first collected three interaction databases to obtain a large aggregated pathway model of human signaling processes. Then they employed a greedy algorithm by initializing a group with an altered gene as a seed, and greedily expanding it with the next best candidate gene until reaching a stop condition to obtain a group with a score. In each step after adding a candidate, the members still have a common downstream target gene that can be reached without traversing any non-member genes. Finally, they performed a permutation test to control the false discovery rate (FDR) in the resulting groups.

They have applied it to 17 different TCGA cancer datasets and identified multiple significantly altered gene groups. They also validated the efficiency of their method by comparing it with existing methods on simulated datasets. According to the principle of this method, all group members are required to be directly linked on the network. Extending this work by allowing non-member linker nodes can detect more distant mutual exclusivity relations. However, this will be challenging because of the extended search space and the reduced statistical power due to multiple hypothesis testing ~\cite{Babur}.

\subsubsection{De Novo Identification Methods}

\begin{figure*}[!t]
\centering
\includegraphics[width=4.8in]{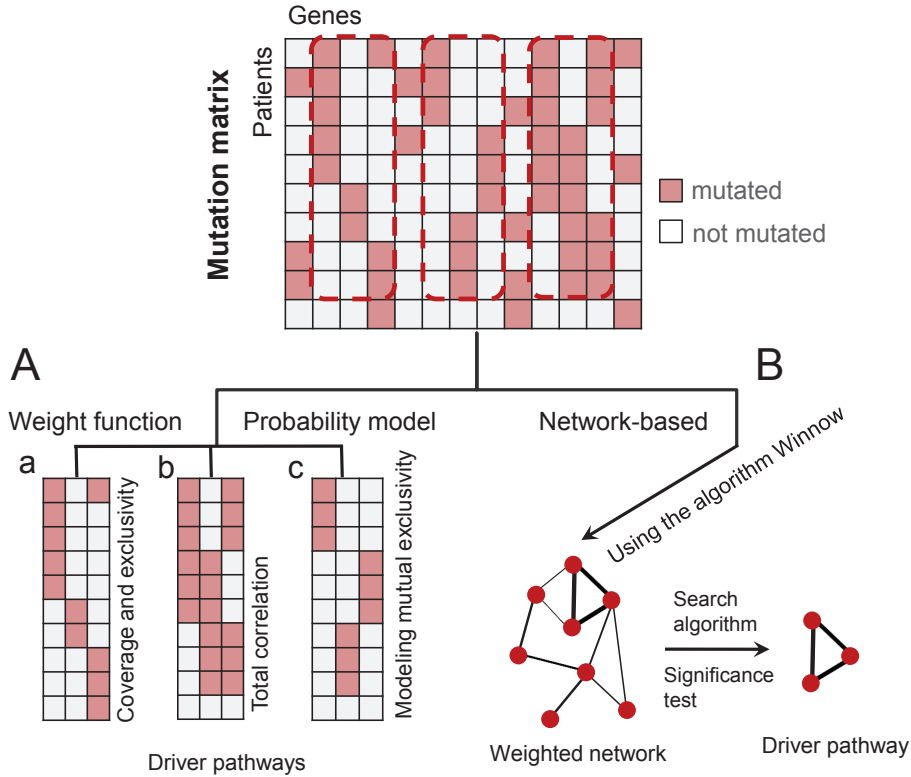}
\caption{A schematic illustration of the methods for \emph{de novo} identifying driver pathways. (A)	Direct extraction methods based on the mutation matrix. (a) Use a weight function (such as $W$) and identify a gene set with high coverage and mutual exclusivity mutations ~\cite{Vandin,Zhao}. (b) Use the weight function `total correlation' and identify a gene set with the so called `exclusive or' pattern ~\cite{Melamed}. (c) Use a probabilistic, generative model of mutual exclusivity and identify the genes with similar mutation ratios ~\cite{Szczurek}. (B) Network-based extraction from the mutation matrix. Build a weighted network by using the algorithm Winnow; then extract core modules using a greedy search algorithm and significance test ~\cite{Miller}.}
\label{fig3}
\end{figure*}

\noindent For these approaches stated above, prior knowledge such as interactions between genes/proteins or known pathways are required. However, this kind of knowledge is far from completeness now. Restricting attention to such information limits the discovery of novel combinations of mutated genes. To identify more driver pathways or gene sets, it would be ideal to assess the significance of recurrent mutations of all possible combinations of genes. But such a \emph{de novo} approach seems implausible because of the huge number of combinations of mutated genes to test. Fortunately, considering mutual exclusivity often possessed by driver mutations, one can use this property as a criterion to \emph{de novo} detect driver gene sets (Fig.~\ref{fig3}).

\vskip3mm
\noindent\textbf{Dendrix}

\noindent Besides the high exclusivity, high coverage is another key characteristic of a driver pathway. In other words, a driver pathway tends to be perturbed in a relatively large number of patients. These two factors could dramatically limit the search space and have been combined to identify \emph{de novo} driver pathways ~\cite{Vandin,Zhao}. Vandin \emph{et al.} ~\cite{Vandin} introduced a weight function $W$ to reward coverage while penalize overlap to get high exclusivity. Specifically, given a binary mutation matrix $A$ with $m$ rows (samples) and $n$ columns (genes), the goal is to find a submatrix $M$ of size $m\times k$ in the mutation matrix $A$ by maximizing the weight function $W$:
\begin{equation}\label{eq:1}
W(M)=|\Gamma (M)|-\omega (M)=2|\Gamma (M)|-\sum_{g\in M}|\Gamma (g)|,
\end{equation}
where $\Gamma (g)=\{i: A_{ig}=1\}$ denotes the set of patients in which gene $g$ is mutated, $\Gamma (M)=\cup_{g\in M} \Gamma (g)$ and $\omega (M)=\sum_{g\in M}|\Gamma (g)|-|\Gamma (M)|$. $|\Gamma (M)|$ measures the coverage of $M$ and $\omega (M)$ measures the coverage overlap of $M$. This was named as the maximum weight submatrix problem. Vandin \emph{et al.} ~\cite{Vandin} introduced a greedy algorithm and a Markov chain Monte Carlo (MCMC) approach (called Dentrix) to solve it. The MCMC approach samples from sets of genes in proportion to their weight $W$ (i.e., the gene sets with high coverage and exclusivity have a higher probability to be chosen). The authors have not paid more attention on the greedy algorithm because of its requirement for a large sample size and the hypothesis of gene independence ~\cite{Vandin}. They have applied the MCMC approach to three cancer mutation datasets and for some pre-assigned gene numbers (i.e., $k$'s) they identified several driver gene sets relating to the key cancer processes including \emph{Rb, p53, mTOR}, and \emph{MAPK} signaling pathways. In most cases, one can used the MCMC approach to get a gene set with a large value of the weight $W$, but a possibility is that MCMC may be trapped in a local solution because of the stochastic search process.

\vskip3mm
\noindent\textbf{MDPFinder}

\noindent Zhao \emph{et al}. ~\cite{Zhao} developed a package MDPFinder (Mutated Driver Pathway Finder) including an exact model and a stochastic search algorithm to find mutated driver pathways. To address the issue of local optimality of Dendrix, Zhao \emph{et al.} first proposed a binary linear programming (BLP) model to exactly solve the maximum weight submatrix problem ~\cite{Zhao}. BLP can be employed to assess the accuracy of other approximate and/or heuristic algorithms. Due to the sparse structure of the mutation data, BLP is much faster than Dendrix. Thus, it can be applied to the analysis of large-scale mutation data sets.

Zhao \emph{et al.} ~\cite{Zhao} also developed a genetic algorithm (GA) to maximize more general and flexible weight functions (including $W$). In other words, it can be applied to a weight function with different format or a new weight function for incorporating other types of data (e.g., gene expression data) to identify more biologically relevant gene sets. The principle of integrating gene expression data comes from that the expression profiles of gene pairs in the same pathway usually have higher correlations than those in different pathways ~\cite{Qiu}. This model can be employed to distinguish the genes that have identical mutation profiles or identify some gene sets with suboptimal score $W$ but with significant biological relevance.

\vskip3mm
\noindent\textbf{ME}

\noindent Cancer genomic alteration data may contain errors such as measurement noise, false mutation calls and their misinterpretation, which can severely bias evaluation and ranking of alteration patterns. Szczurek \emph{et al.} ~\cite{Szczurek} developed probabilistic models to analyze cancer alteration data and then to detect gene sets with mutually exclusive (ME) mutations. They introduced two models: the first one is a probabilistic, generative model of mutually exclusive patterns, which considers observation errors; the second one is a null model assuming independent alterations of genes. Comparing the first model to the second one, a statistical test of mutual exclusivity is derived. The generative model assumes that genes in a module have equal chance to be altered. Thus, the detected modules may tend to contain genes with similar alteration ratios ~\cite{Babur}. This may be improved by considering the real mutation distribution.

\vskip3mm
\noindent\textbf{RME}

\noindent Miller \emph{et al.} ~\cite{Miller} developed the RME algorithm to detect functional modules in tumors based on the patterns of recurrent and mutually exclusive (RME) aberrations. First, they filtered the mutation matrix to get genes satisfying a pre-specified mutation frequency. Next, they used an online-learning linear threshold algorithm called Winnow to score each gene pair by exclusivity and created a weighted gene network. Then, they obtained a set of candidate RME modules by a greedy local combinatorial search. Finally, they employed an algorithmic significance test to evaluate the significance of the RME patterns and thus determined the significant RME modules. The RME algorithm only considers genes mutated with relatively high frequency which may limit its effectiveness in identifying rare driver mutations.

\vskip3mm
\noindent\textbf{iMCMC}

\noindent Zhang \emph{et al}. developed a network-based method to identify Mutated Core Modules in Cancer (iMCMC) by integrating somatic mutations, CNAs, and gene expressions ~\cite{Zhang_2013}. They have considered four gene features in the driver pathways or core modules: high coverage, mutual exclusivity, strong influence of a gene's mutation on other genes and high correlation of gene expressions. Specifically, they first obtained a mutation matrix by merging somatic mutations and CNAs, and constructed a weighted mutation network where the vertex weight corresponds to gene coverage and the edge weight corresponds to the mutual exclusivity between gene pairs. They also generated a weighted expression network from the expression matrix where the vertex and edge weights correspond to the influence of a gene mutation on other genes and the Pearson correlation of gene expressions, respectively. Then they obtained an integrative network by further combining these two networks, and identified the most coherent subnetworks using an optimization model. Finally, they extracted the core modules for tumors with both significance and exclusivity tests. They applied iMCMC to real data to demonstrate that it can identify several mutated core modules involved in known carcinogenesis pathways.

\vskip3mm
\noindent\textbf{SODP}

\noindent We have to note that genes with mutually exclusive mutations alone may be not sufficient to indicate they are in a common pathway. Recently, Lu \emph{et al.} ~\cite{Lu2015} developed a signal-oriented framework for discovering driver pathways (named as SODP) by integrating genomic alteration data and transcriptomic data from TCGA project. First, they identified the perturbed cellular signals by grouping the differentially expressed genes into functional groups summarized by their Gene Ontology (GO) annotation. Next, they constructed a bipartite graph consisting of tumors and genes and searched for a densely connected subgraph. Finally, they designed an exact algorithm to identify a set of mutually exclusive gene alteration events which carries strong information with respect to the perturbed signals among the detected tumors. A main concern about SODP is that the association between the detected gene alterations and the perturbed signals may not have a causal relationship which needs further experimental validation ~\cite{Lu2015}.

\subsection{Identification of Cooperative Driver Pathways}
All the above studies have focused on the identification of individual driver pathways or core modules, where each time a single group of genes with certain properties is identified. However, cancer is a complex disease and multiple pathways with mutations are generally required for carcinogenesis ~\cite{Hanahan_11}. In fact, it has been recognized that pathways often function cooperatively in cancer initiation and progression ~\cite{Yeang,Cui,Klijn}. Besides the combinatorial patterns with co-occurring mutations identified by Yeang \emph{et al.} ~\cite{Yeang}, some other examples involved in lung squamous cell carcinoma and GBM can be found in ~\cite{Justilien} and ~\cite{Gu2013}, respectively.

\begin{figure*}[!t]
\centering
\includegraphics[width=4.8in]{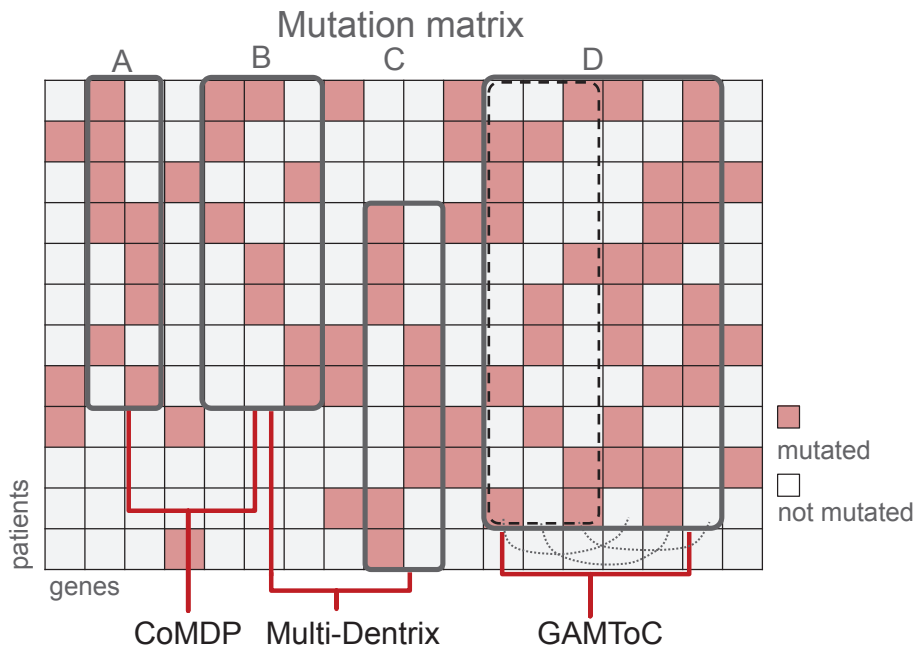}
\caption{A schematic illustration of the methods for the identification of cooperative driver pathways. Given the mutation matrix, CoMDP tends to identify gene sets A and B, each of which has high mutual exclusivity. The co-occurrence of A and B implies likely cooperation between these two pathways ~\cite{Zhang_2014}. Multi-Dentrix may identify gene sets B and C, whose mutations are not necessarily co-occurring although each of them also has high mutual exclusivity ~\cite{Leiserson_13}. GAMToC may identify a gene set with complex mutational patterns (the gene set D). The first three genes are mutually exclusive, and the others are their respective inversion ~\cite{Melamed}.}
\label{fig4}
\end{figure*}

Thus, investigating the complex collaboration among different biological pathways and functional modules can shed new lights on the understanding of the molecular mechanisms of cancer formation and progression. In previous studies, incomplete prior knowledge on pathways and/or interaction networks were utilized to determine whether two or more pathways or modules are simultaneously perturbed in the same samples ~\cite{Yeang,Cui,Gu2013}. Recently, several \emph{de novo} approaches have been developed to discover collaborative pathways playing driver roles in cancer initiation and development ~\cite{Zhang_2014,Leiserson_13,Melamed} (Fig.~\ref{fig4}).

\vskip3mm
\noindent\textbf{CoMDP}

\noindent To investigate the collaboration among different pathways, a natural way is to investigate whether they are almost simultaneously mutated in a large cohort of patients. Zhang \emph{et al.} ~\cite{Zhang_2014} introduced a weight function $H$ to investigate such co-occurring mutation patterns. They further developed a mathematical programming model to maximize $H$ to \emph{de novo} discover co-occurring mutated driver pathways (CoMDP) in cancer. Specifically, for a mutation matrix $A$, they considered two submatrices $M$ and $N$ (which correspond to two gene sets or pathways). Given the coverage $\Gamma (M)$ and $\Gamma (N)$ of the two gene sets, they defined two quantities: the common coverage $c(M,N)=|\Gamma (M)\bigcap \Gamma (N)|$, and the union coverage $b(M,N)=|\Gamma (M)\bigcup \Gamma (N)|$. They further defined the non-shared coverage $d(M,N)=b(M,N)-c(M,N)$, which describes the extent of the mutation co-occurrence between the two gene sets: the smaller the value $d$, the larger the co-occurrence is. As stated in Eq.~(\ref{eq:1}), $\omega (M)$ and $\omega (N)$ reflect the exclusivity of $M$ and $N$ respectively. They introduced the weight function $H$ as follows:
\begin{equation}\label{eq:2}
H(M,N)=c(M,N)-d(M,N)-\omega (M)-\omega (N).
\end{equation}
Maximizing $H$ implies the maximization of two types of characteristics simultaneously: (1) the maximization of the weight $W$ for each individual pathway (i.e., high coverage and high exclusivity; this can be realized by comprehensively considering all the four terms on the right side of Eq.~(\ref{eq:2})); and (2) the maximization of the inter-overlap between the pathway pair (guaranteed by the first two terms on the right side of Eq.~(\ref{eq:2})). Moreover, CoMDP is an exact method where the optimal set of pathways is obtained using an efficient algorithm due to the sparse data structure. It does not require any prior knowledge besides mutation profiles. They have demonstrated that CoMDP can get the exact solution of gene sets with significant co-occurring mutations using simulation data. They applied CoMDP to several real biological data and discovered co-occurring driver pathways involved in several key biological processes such as cell survival and protein synthesis. Furthermore, they also proposed a modified form (named mod\_CoMDP) to model the situations that a certain pathway has been previously proven to play important roles in some cancers and one wants to know whether there are other pathways with cooperative effects in carcinogenesis ~\cite{Zhang_2014}.

\vskip3mm
\noindent\textbf{Multi-Dendrix}

\noindent Leiserson \emph{et al.} ~\cite{Leiserson_13} generalized Dentrix ~\cite{Vandin} and proposed an approach, called Multi-Dendrix, to simultaneously identify multiple driver pathways in cancer. Multi-Dendrix was designed to optimize the weight function which is a sum of $r$ quantities for the weight of $r$ pathways. Each of the $r$ quantities corresponds to $W$ in Eq.~(\ref{eq:1}), so each of the gene sets detected by Multi-Dendrix has high coverage and high exclusivity. The authors also used a binary linear programming to solve this problem, when $r=1$, which is equivalent to the one proposed in MDPFinder ~\cite{Zhao}. Applying Multi-Dendrix to somatic mutations from GBM, breast cancer, and lung cancer samples, the authors identified multiple sets of genes involved in known important pathways for carcinogenesis. There is no considering the relationship among the mutations of the gene sets, so the identified multiple pathways by Multi-Dendrix are not necessarily co-occurring ~\cite{Zhang_2014}.

\vskip3mm
\noindent\textbf{GAMToC}

\noindent Recently, Melamed \emph{et al.} ~\cite{Melamed} introduced an information theoretic method, called GAMToC, to identify combinations of genomic alterations in cancer. GAMToC adopts the total correlation (i.e., mutual information) to measure the difference between the joint uncertainty (or entropy) of a set of variables (genes) and their individual uncertainties. The key assumption is that the high total correlation suggests a joint relationship among individual genes. Whereas, when there is no joint relationship between the variables, the difference will be zero.

A main property of GAMToC is that it detects a gene set with jointly related mutation patterns each time. In such a gene set, some genes may demonstrate high mutual exclusivity, and some other genes may show high co-occurrence. An example is an `exclusive or' triplet of genes where lesion of any two of the genes is enough to change a phenotype, and the third adds no further advantage. Although the genes display no mutual exclusivity or co-occurrence characteristic, the total correlation of this three-gene pattern is highly significant (Fig.~\ref{fig3}A-b). The patterns detected by GAMToC may be very complicated (Fig.~\ref{fig4}), indicating a complex cooperation among multiple biological pathways, which needs further investigation.

\vskip3mm
\noindent\textbf{LM}

\noindent Although co-occurring or mutual exclusive gene alterations have been explored in cancer, the understanding of molecular mechanisms in cancer initiation and progression remains challenging. A basic problem is that pathways involved in carcinogenesis are complex and interconnected without clear boundary. Recently, Remy \emph{et al.} developed a logical model (LM) to explain mutually exclusive and co-occurring genetic alterations (including somatic mutations, CNAs) in bladder carcinogenesis ~\cite{Remy_2015}. First, by performing literature search and data mining of four independent bladder cancer datasets, they identified nine patterns of co-occurrence and mutual exclusivity in genetic alterations which are involved in growth factor signaling pathways, cell cycle and apoptosis. Next, they organized the interactions between these genes into an influence network based on literature analysis. Then they introduced a logical model and analyzed it with the GINsim software. They validated this model using a published mutant mice data. Moreover, it was used to study the co-occurring and mutual exclusivity patterns, suggesting this model allows one to formulate predictions about conditions where combining genetic alterations benefits carcinogenesis.

Indeed, by combining prior knowledge and mathematical modeling, LM can shed some lights on the mechanisms leading to carcinogenesis. However, the model only includes simplified representations of pathways, so in some cases the interpretation of the results from the model analysis must be done with further validation ~\cite{Remy_2015}.

\subsection{Pan-cancer Scale Analysis}
With the accumulation of a large number of human cancer genomics data, the TCGA pan-cancer project surveyed multi-platform aberration data in cancer samples from thousands of cancer patients among 12 cancer types ~\cite{Weinstein_13}. The rich data provide a major opportunity to systematically examine the similarities and differences among multiple cancer types ~\cite{Weinstein_13,Liu2014}. Recently, several studies investigated the mutual exclusivity and/or co-occurrence characteristics across large TCGA pan-cancer mutation datasets ~\cite{Leiserson_15,Kim_15}.

\vskip3mm
\noindent\textbf{HotNet2}

\noindent As mentioned earlier, HotNet2 ~\cite{Leiserson_15} was designed to identify pathways and protein complexes based on pan-cancer network analysis. HotNet2 employs a directed heat diffusion model to detect mutated subnetworks without considering within-subnetwork mutual exclusivity and across-subnetwork co-occurrence directly. Many pairs of detected subnetworks exhibit significant co-occurrence across the pan-cancer cohort or in individual cancer types. Comparatively, mutual exclusivity is typically possessed by the gene mutations within a pathway but not across pathways. These observations are consistent with previous studies ~\cite{Yeang,Vandin}, and also support the hypothesis that cancer cells harbor multiple driver mutations that perturb multiple biological functions ~\cite{Hanahan_11}.

\vskip3mm
\noindent\textbf{MEMCover}

\noindent As we have surveyed, mutual exclusivity has been adopted for identifying driver pathways as a primary property of gene mutations in cancer ~\cite{Vandin,Zhao,Ciriello,Zhang_2014}. However, recent pan-cancer studies ~\cite{Kandoth,Szczurek} found that some mutually exclusive genes are cancer type-specific. Thus, it is necessary to recognize different mutual exclusivity classes in pan-cancer analysis for driver pathway identification.

Kim \emph{et al.} ~\cite{Kim_15} classified mutual exclusivity into three classes and carefully studied their properties. The first class is `within tissue type exclusivity' which means it is observed only in one cancer type. The second class is called `across tissue type exclusivity', which is observed in more than one tissue types. The last one is `between tissue type exclusivity', which is observed between tissue-specific genes. The first two classes of mutually exclusive gene pairs have been demonstrated to have a high possibility to be interacted in the functional networks than those in the last class. The method MEMCover was designed by combining across tissue type exclusivity with interaction data to uncover pan-cancer dysregulated pathways. The identified subnetworks not only contain previously known pan-cancer dysregulated modules but also reveal novel ones whose across cancer role has not been explored well before. It is notable that the selection of subnetworks in MEMCover is guided by mutual exclusivity, interaction network connectivity and sample coverage. Thus, the subnetworks detected by MEMCover do not necessarily contain mutually exclusive pairs ~\cite{Kim_15}.

\section{Conclusion}

The patterns of mutually exclusive genomic alterations provide important clue for understanding cancer initiation and progression, and have widely been used to identify driver pathways or modules in cancer in recent years. In fact, besides the popular interpretation about mutual exclusivity, i.e., alteration to a second gene within the same pathway offers no further selective advantage, there is another explanation on it, which says that the second alteration within the same pathway actually leads to a disadvantage for the cell, even results in cell death. This is referred to as synthetic lethality ~\cite{Szczurek_13,Wang_X}, which may provide an alternative strategy for cancer treatment ~\cite{Jerby-Arnon}. Although these two hypotheses cannot be systematically distinguished based on genomic data alone, the observed mutual exclusivity provides evidence that the altered genes are functionally linked, and most likely linked in a common pathway or biological process.

Recent studies have developed a large number of algorithms and models for the identification of driver pathways or core modules ~\cite{Vandin_11,Cerami_10,Ciriello,Vandin,Zhao,Miller,Szczurek,Zhang_2013}. These methods have been applied to real data and revealed key biological pathways such as \emph{p53, Rb} and \emph{PI(3)K} pathways. Previous studies have provided evidence that carcinogenesis is a complex process and the malignant transformation from a normal cell to a tumor may be a highly cooperative procedure involving synergy between pathways. The methods for the identification of cooperative driver pathways such as CoMDP ~\cite{Zhang_2014}, Multi-Dentrix ~\cite{Leiserson_13} and GAMToC ~\cite{Melamed} may be valuable to advance such analysis. CoMDP can identify co-occurring mutated driver pathways, while Multi-Dentrix can identify multiple pathways simultaneously without guaranteeing their co-occurrence in a cohort of patients. GAMToC can detect a set of genes which may contain multiple pathways, wherein some mutations are exclusive, some are co-occurring, and some have the `exclusive or' relationship or others. However, how various cellular and physiological processes are coordinately altered during the initiation and progression of cancer is still a major challenge and need more deep investigations in the future.

Here, we review the recent methods for identifying driver genes, driver pathways in cancer using genomic alteration data. However, large-scale cancer genomics projects provide huge numbers of multiple platform data including gene expression, DNA methylation, microRNA expression, protein expression, and clinical data ~\cite{TCGA_08,TCGA_11}. Different kinds of data can provide different information for cancer research. How to integrate diverse data into one model or framework to investigate cancer initiation and progression is an important challenge. There have been a lot of efforts in this direction. Besides the methods mentioned above ~\cite{Zhang_2013,Lu2015}, other approaches include PARADIGM ~\cite{Vaske} which infers patient-specific pathway activities by using multi-dimensional cancer genomics data, CONEXIC ~\cite{Akavia} which integrates CNAs and gene expression data to identify driver mutations and the processes that they influence, and DriverNet ~\cite{Bashashati} which was designed to relate genomic aberrations to disrupted transcriptional patterns through molecular interaction networks to identify driver genes in cancer. Several integrative methods to decipher coherent  patterns have also been developed ~\cite{Zhang_12,Li_12,Mo_13,Masica,TCGA_13,Cheng,Kristensen}, which may help to reveal clinically relevant characteristics. More efforts are needed to comprehensively integrate diverse cancer genomics data to uncover complex mechanisms underlying oncogenesis.

Cancer displays large heterogeneity, but different types of cancers may possess commonalities ~\cite{Weinstein_13}. The rich cancer genomics data provide a major opportunity to develop an integrated picture of commonalities and differences across tumor lineages, which is the main goal of the TCGA pan-cancer project ~\cite{Weinstein_13,Liu2014,Liu2015}. Although some comparative studies among multiple cancer types have been developed in recent years ~\cite{Leiserson_15,Kim_15,Ciriello_13,Lawrence_14,Liu2015}, more further explorations are needed and more comprehensive results are expected. Analysis of the molecular aberrations and their functional roles across tumor types will be helpful to extend effective therapies in one cancer type to others with a similar genomic profile.

We also realize that along with more and more computational models and algorithms are developed to identify driver pathways in cancer, a reasonable assessment or comparison of the performance of the approaches is of pressing need. For driver gene identification methods, a comparative strategy may be ranking the detected genes first by some scores, and then investigating how many genes in the top are in COSMIC or other cancer related database. But for driver pathway approaches, a main issue is that pathways involved in carcinogenesis are complex and interconnected without clear boundary, so it is difficult to give a uniform benchmark to evaluate the performance of these methods. Even so, more efforts are expected to solve this challenge as soon as possible because reasonable evaluation will benefit one to develop more effective methods to decipher the pathogenesis underlying cancer and thus will help understand the molecular mechanism of carcinogenesis and make effective personalized treatments for cancer patients.

\section*{Acknowledgments}
Shihua Zhang is the corresponding author of this paper. This work was supported by the National Natural Science Foundation of China, No. 61379092, 61422309 and 11131009, the Strategic Priority Research Program of the Chinese Academy of Sciences (CAS) (XDB13040600), the Outstanding Young Scientist Program of CAS and the Key Laboratory of Random Complex Structures and Data Science, CAS (No. 2008DP173182).

\ifCLASSOPTIONcaptionsoff
  \newpage
\fi


\begin{thebibliography}{1}

\bibitem{TCGA_08}
The Cancer Genome Atlas Research Network, ``Comprehensive genomic characterization defines human glioblastoma genes and core pathways," \emph{Nature}, vol. 455, pp. 1061-1068, 2008.

\bibitem{TCGA_11}
The Cancer Genome Atlas Research Network, ``Integrated genomic analyses of ovarian carcinoma," \emph{Nature}, vol. 474, pp. 609-615, 2011.

\bibitem{ICGC}
International Cancer Genome Consortium, ``International network of cancer genome projects," \emph{Nature}, vol. 464, pp. 993-998, 2010.

\bibitem{Barretina}
J.~Barretina, G.~Caponigro, N.~Stransky, K.~Venkatesan, A.~A. Margolin, S.~Kim, \emph{et al.}, ``The Cancer Cell Line Encyclopedia enables predictive modelling of anticancer drug sensitivity," \emph{Nature}, vol. 483, pp. 603-607, 2012.

\bibitem{Mullighan}
C.~G. Mullighan, X.~Su, J.~Zhang, I.~Radtke, L.~A. Phillips, C.~B. Miller, \emph{et al.}, ``Deletion of IKZF1 and Prognosis in Acute Lymphoblastic Leukemia," \emph{N Engl J Med}, vol. 360, pp. 470-480, 2009.

\bibitem{Jiang_14}
Y.~Jiang, D.~Redmond, K.~Nie, K.~W. Eng, T.~Clozel, P.~Martin, \emph{et al.}, ``Deep-sequencing reveals clonal evolution patterns and mutation events associated with relapse in B-cell lymphomas," \emph{Genome Biol}, vol. 15, 432, 2014.

\bibitem{Hofree}
M.~Hofree, J.~P. Shen, H.~Carter, A.~Gross, and T.~Ideker, ``Network-based stratification of tumor mutations," \emph{Nat Methods}, vol. 10, pp. 1108-1115, 2013.

\bibitem{Hansen}
K.~D. Hansen, W.~Timp, H.~C. Bravo, S.~Sabunciyan, B.~Langmead, O.~G. McDonald, \emph{et al.}, ``Increased methylation variation in epigenetic domains across cancer types," \emph{Nat Genet}, vol. 43, pp. 768-775, 2011.

\bibitem{Jiang_15}
P.~Jiang, M.~L. Freedman, J.~S. Liu, and X.~S. Liu, ``Inference of transcriptional regulation in cancers," \emph{Proc Natl Acad Sci}, vol. 112, pp. 7731-7736, 2015.

\bibitem{Yates}
L.~R. Yates and P.~J. Campbell, ``Evolution of the cancer genome," \emph{Nat Rev Genet}, vol. 13, pp. 795-806, 2012.

\bibitem{Sun}
Y.~Sun, J.~Yao, N.~J. Nowak, and S.~Goodison, ``Cancer progression modeling using static sample data," \emph{Genome Biol}, vol. 15, 440, 2014.

\bibitem{Wang_J}
J.~Wang, H.~Khiabanian, D.~Rossi, G.~Fabbri, V.~Gattei, F.~Forconi, \emph{et al.}, ``Tumor evolutionary directed graphs and the history of chronic lymphocytic leukemia," \emph{Elife}, 2014; vol. 3, e02869, 2014.

\bibitem{Lu_05}
J.~Lu, G.~Getz, E.~A. Miska, E.~Alvarez-Saavedra, J.~Lamb, D.~Peck, \emph{et al.}, ``MicroRNA expression profiles classify human cancers," \emph{Nature}, vol. 435, pp. 834-838, 2005.

\bibitem{Reis-Filho}
J.~S. Reis-Filho and L.~Pusztai, ``Gene expression profiling in breast cancer: classification, prognostication, and prediction," \emph{Lancet}, vol. 378, pp. 1812-1823, 2011.

\bibitem{Liu}
Z.~Liu, X.~S. Zhang, and S.~Zhang, ``Breast tumor subgroups reveal diverse clinical predictive power," \emph{Sci Rep}, 2014; vol. 4, 4002, 2014.

\bibitem{Ding_08}
L.~Ding, G.~Getz, D.~A. Wheeler, E.~R. Mardis, M.~D. McLellan, K.~Cibulskis, \emph{et al.}, ``Somatic mutations affect key pathways in lung adenocarcinoma," \emph{Nature}, vol. 455, pp. 1069-1075, 2008.

\bibitem{Youn_11}
A.~Youn and R.~Simon, ``Identifying cancer driver genes in tumor genome sequencing studies," \emph{Bioinformatics}, vol. 27, pp. 175-181, 2011.

\bibitem{Lawrence}
M.~S. Lawrence, P.~Stojanov, P.~Polak, G.~V. Kryukov, K.~Cibulskis, A.~Sivachenko, \emph{et al.}, ``Mutational heterogeneity in cancer and the search for new cancer-associated genes," \emph{Nature}, vol. 499, pp. 214-218, 2013.

\bibitem{Sjoblom}
T.~Sj$\ddot{o}$blom, S.~Jones, L.~D. Wood, D.~W. Parsons, J.~Lin, T.~D. Barber, \emph{et al.}, ``The consensus coding sequences of human breast and colorectal cancers," \emph{Science}, vol. 314, pp. 268-274, 2006.

\bibitem{Stamatoyannopoulos}
J.~A. Stamatoyannopoulos, I.~Adzhubei, R.~E. Thurman, G.~V. Kryukov, S.~M. Mirkin, and S.~R. Sunyaev, ``Human mutation rate associated with DNA replication timing," \emph{Nat Genet}, vol. 41, pp. 393-395, 2009.

\bibitem{Chen_2010}
C.~L. Chen, A.~Rappailles, L.~Duquenne, M.~Huvet, G.~Guilbaud, L.~Farinelli, \emph{et al.}, ``Impact of replication timing on non-CpG and CpG substitution rates in mammalian genomes," \emph{Genome Res}, vol. 20, pp. 447-457, 2010.

\bibitem{Ng_PC}
P.~C. Ng and S.~Henikoff, ``Predicting deleterious amino acid substitutions," \emph{Genome Res}, vol. 11, pp. 863-874, 2001.

\bibitem{Adzhubei}
I.~A. Adzhubei, S.~Schmidt, L.~Peshkin, V.~E. Ramensky, A.~Gerasimova, P.~Bork, \emph{et al.}, ``A method and server for predicting damaging missense mutations," \emph{Nat Methods}, vol. 7, pp. 248-249, 2010.

\bibitem{Reva}
B.~Reva, Y.~Antipin, and C.~Sander, ``Predicting the functional impact of protein mutations: Application to cancer genomics," \emph{Nucleic Acids Res}, vol. 39, e118, 2011.

\bibitem{Dees}
N.~D. Dees, Q.~Zhang, C.~Kandoth, M.~C. Wendl, W.~Schierding, D.~C. Koboldt, \emph{et al.}, ``MuSiC: identifying mutational significance in cancer genomes," \emph{Genome Res}, vol. 22, pp. 1589-1598, 2012.

\bibitem{Vogelstein_13}
B.~Vogelstein, N.~Papadopoulos, V.~E. Velculescu, S.~Zhou, L.~A. Diaz Jr., and K.~W. Kinzler, ``Cancer genome landscapes," \emph{Science}, vol. 339, pp. 1546-1558, 2013.

\bibitem{Tamborero}
D.~Tamborero, A.~Gonzalez-Perez, and N.~Lopez-Bigas, ``Oncodriveclust: exploiting the positional clustering of somatic mutations to identify cancer genes," \emph{Bioinformatics}, vol. 29, pp. 2238-2244, 2013.

\bibitem{Korthauer}
K.~D. Korthauer and C.~Kendziorski, ``MADGiC: a model-based approach for identifying driver genes in cancer," \emph{Bioinformatics}, vol. 31, pp. 1526-1535, 2015.

\bibitem{Nik-Zainal}
S.~Nik-Zainal, P.~Van Loo, D.~C. Wedge, L.~B. Alexandrov, C.~D. Greenman, K.~W. Lau, \emph{et al.}, ``The life history of 21 breast cancers," \emph{Cell}, vol. 149, pp. 994-1007, 2012.

\bibitem{Vogelstein_04}
B.~Vogelstein and K.~W. Kinzler, ``Cancer genes and the pathways they control," \emph{Nat Med}, vol. 10, pp. 789-799, 2004.

\bibitem{Ding}
L.~Ding, G.~Getz, D.~A. Wheeler, E.~R. Mardis, M.~D. McLellan, K.~Cibulskis, \emph{et al.}, ``Somatic mutations affect key pathways in lung adenocarcinoma," \emph{Nature}, vol. 455, pp. 1069-1075, 2008.

\bibitem{Hanahan_00}
D.~Hanahan and R.~A. Weinberg, ``The hallmarks of cancer," \emph{Cell}, vol. 100, pp. 57-70, 2000.

\bibitem{Hanahan_11}
D.~Hanahan and R.~A. Weinberg, ``Hallmarks of cancer: the next generation," \emph{Cell}, vol. 144, pp. 646-674, 2011.

\bibitem{Subramanian}
A.~Subramanian, P.~Tamayo, V.~K. Mootha, S.~Mukherjee, B.~L. Ebert, M.~A. Gillette, \emph{et al.}, ``Gene set enrichment analysis: a knowledge-based approach for interpreting genome-wide expression profiles," \emph{Proc Natl Acad Sci}, vol. 102, pp. 15545-15550, 2005.

\bibitem{Lin_07}
J.~Lin, C.~M. Gan, X.~Zhang, S.~Jones, T.~Sj$\ddot{o}$blom, L.~D. Wood, \emph{et al.}, ``A multidimensional analysis of genes mutated in breast and colorectal cancers," \emph{Genome Res}, vol. 17, pp. 1304-1318, 2007.

\bibitem{Boca}
S.~M. Boca, K.~W. Kinzler, V.~E. Velculescu, B.~Vogelstein, and G.~Parmigiani, ``Patient-oriented gene set analysis for cancer mutation data," \emph{Genome Biol}, vol. 11, R112, 2010.

\bibitem{Wendl}
M.~C. Wendl, J.~W. Wallis, L.~Lin, C.~Kandoth, E.~R. Mardis, R.~K. Wilson, \emph{et al.}, ``PathScan: a tool for discerning mutational significance in groups of putative cancer genes," \emph{Bioinformatics}, vol. 27, pp. 1595-1602, 2011.

\bibitem{Garraway}
L.~A. Garraway and E.~S. Lander, ``Lessons from the cancer genome," \emph{Cell}, vol. 153, pp. 17-37, 2013.

\bibitem{Wu_10}
G.~Wu, X.~Feng, and L.~Stein, ``A human functional protein interaction network and its application to cancer data analysis," \emph{Genome Biol}, vol. 11, R53, 2010.

\bibitem{Vandin_11}
F.~Vandin, E.~Upfal, and B.~J. Raphael, ``Algorithms for detecting significantly mutated pathways in cancer," \emph{J Comput Biol}, vol. 18, pp. 507-522, 2011.

\bibitem{Leiserson_15}
M.~D.~M. Leiserson, F.~Vandin, H.~T. Wu, J.~R. Dobson, J.~V. Eldridge, J.~L. Thomas, \emph{et al.}, ``Pan-cancer network analysis identifies combinations of rare somatic mutations across pathways and protein complexes," \emph{Nat Genet}, vol. 47, pp. 106-114, 2015.

\bibitem{Cerami_10}
E.~Cerami, E.~Demir, N.~Schultz, B.~S. Taylor, and C.~Sander, ``Automated network analysis identifies core pathways in glioblastoma," \emph{PLoS One}, vol. 5, e8918, 2010.

\bibitem{Ciriello}
G.~Ciriello, E.~Cerami, C.~Sander, and N.~Schultz, ``Mutual exclusivity analysis identifies oncogenic network modules," \emph{Genome Res}, vol. 22, pp. 398-406, 2012.

\bibitem{Yeang}
C.~H. Yeang, F.~McCormick, and A.~Levine, ``Combinatorial patterns of somatic gene mutations in cancer," \emph{FASEB J}, vol. 22, pp. 2605-2622, 2008.

\bibitem{Vogelstein_88}
B.~Vogelstein, E.~R. Fearon, S.~R. Hamilton, S.~E. Kern, A.~C. Preisinger, M.~Leppert, \emph{et al.}, ``Genetic alterations during colorectal-tumor development," \emph{N Engl J Med}, vol. 319, pp. 525-532, 1988.

\bibitem{Cahill}
D.~P. Cahill, K.~W. Kinzler, B.~Vogelstein, and C.~Lengauer, ``Genetic instability and darwinian selection in tumours," \emph{Trends Cell Biol}, vol. 9, pp. M57-M60, 1999.

\bibitem{Mermel}
C.~H. Mermel, S.~E. Schumacher, B.~Hill, M.~L. Meyerson, R.~Beroukhim, and G.~Getz, ``GISTIC2.0 facilitates sensitive and confident localization of the targets of focal somatic copy-number alteration in human cancers," \emph{Genome Biol}, vol. 12, R41, 2011.

\bibitem{Babur}
$\ddot{O}$.~Babur, M.~G$\ddot{o}$nen, B.~A. Aksoy, N.~Schultz, G.~Ciriello, C.~Sander, \emph{et al.}, ``Systematic identification of cancer driving signaling pathways based on mutual exclusivity of genomic alterations," \emph{Genome Biol}, vol. 16, 45, 2015.

\bibitem{Vandin}
F.~Vandin, E.~Upfal, and B.~J. Raphael, ``\emph{De novo} discovery of mutated driver pathways in cancer," \emph{Genome Res}, vol. 22, pp. 375-385, 2012.

\bibitem{Zhao}
J.~Zhao, S.~Zhang, L.~Y. Wu, and X.~S. Zhang, ``Efficient methods for identifying mutated driver pathways in cancer," \emph{Bioinformatics}, vol. 28, pp. 2940-2947, 2012.

\bibitem{Qiu}
Y.~Q. Qiu, S.~Zhang, X.~S. Zhang, and L.~Chen, ``Detecting disease associated modules and prioritizing active genes based on high throughput data," \emph{BMC Bioinformatics}, vol. 11, 26, 2010.

\bibitem{Szczurek}
E.~Szczurek and N.~Beerenwinkel, ``Modeling mutual exclusivity of cancer mutations," \emph{PLoS Comput Biol}, vol. 10, e1003503, 2014.

\bibitem{Miller}
C.~A. Miller, S.~H. Settle, E.~P. Sulman, K.~D. Aldape, and A.~Milosavljevic, ``Discovering functional modules by identifying recurrent and mutually exclusive mutational patterns in tumors," \emph{BMC Med Genomics}, vol. 4, 34, 2011.

\bibitem{Zhang_2013}
J.~Zhang, S.~Zhang, Y.~Wang, and X.~S. Zhang, ``Identification of mutated core cancer modules by integrating somatic mutation, copy number variation, and gene expression data," \emph{BMC Syst Biol}, vol. 7, S4, 2013.

\bibitem{Lu2015}
S.~Lu, K.~N. Lu, S.~Y. Cheng, B.~Hu, X.~Ma, and N.~Nystrom, ``Identifying driver genomic alterations in cancers by searching minimum-weight, mutually exclusive sets," \emph{PLoS Comput Biol}, vol. 11, e1004257, 2015.

\bibitem{Cui}
Q.~Cui, Y.~Ma, M.~Jaramillo, H.~Bari, A.~Awan, S.~Yang, \emph{et al.}, ``A map of human cancer signaling," \emph{Mol Syst Biol}, vol. 3, 152, 2007.

\bibitem{Klijn}
C.~Klijn, J.~Bot, D.~J. Adams, M.~Reinders, L.~Wessels, and J.~Jonkers, ``Identification of networks of co-occurring, tumor-related DNA copy number changes using a genome-wide scoring approach," \emph{PLoS Comput Biol}, vol. 6, e1000631, 2010.

\bibitem{Justilien}
V.~Justilien, M.~P. Walsh, S.~A. Ali, E.~A. Thompson, N.~R. Murray, and A.~P. Fields, ``The \emph{PRKCI} and \emph{SOX2} oncogenes are coamplified and cooperate to activate Hedgehog signaling in lung squamous cell carcinoma," \emph{Cancer Cell}, vol. 25, pp. 139-151, 2014.

\bibitem{Gu2013}
Y.~Gu, H.~Wang, Y.~Qin, Y.~Zhang, W.~Zhao, L.~Qi, \emph{et al.}, ``Network analysis of genomic alteration profiles reveals co-altered functional modules and driver genes for glioblastoma," \emph{Mol Biosyst}, vol. 9, pp. 467-477, 2013.

\bibitem{Zhang_2014}
J.~Zhang, L.~Y. Wu, X.~S. Zhang, and S.~Zhang, ``Discovery of co-occurring driver pathways in cancer," \emph{BMC Bioinformatics}, vol. 15, 271, 2014.

\bibitem{Leiserson_13}
M.~D.~M. Leiserson, D.~Blokh, R.~Sharan, and B.~J. Raphael, ``Simultaneous identification of multiple driver pathways in cancer," \emph{PLoS Comput Biol}, vol. 9, e1003054, 2013.

\bibitem{Melamed}
R.~D. Melamed, J.~Wang, A.~Iavarone, and R.~Rabadan, ``An information theoretic method to identify combinations of genomic alterations that promote glioblastoma," \emph{J Mol Cell Biol}, vol. 7, pp. 203-213, 2015.

\bibitem{Remy_2015}
E.~Remy, S.~Rebouissou, C.~Chaouiya, A.~Zinovyev, F.~Radvanyi, and I.~Calzone, ``A modelling approach to explain mutually exclusive and co-occurring genetic alterations in bladder tumorigenesis," \emph{Cancer Res}, vol. 75, pp. 4042-4052, 2015.

\bibitem{Weinstein_13}
The Cancer Genome Atlas Research Network, \emph{et al.}, ``The Cancer Genome Atlas Pan-Cancer analysis project," \emph{Nat Genet}, vol. 45, pp. 1113-1120, 2013.

\bibitem{Liu2014}
Z.~Liu Z and S.~Zhang, ``Toward a systematic understanding of cancers: a survey of the pan-cancer study," \emph{Front Genet}, vol. 5, 194, 2014.

\bibitem{Kim_15}
Y.~A. Kim, D.~Y. Cho, P.~Dao, and T.~M. Przytycka, ``MEMCover: Integrated analysis of mutual exclusivity and functional network reveals dysregulated pathways across multiple cancer types," \emph{Bioinformatics}, vol. 31, pp. i284-i292, 2015.

\bibitem{Kandoth}
C.~Kandoth, M.~D. McLellan, F.~Vandin, K.~Ye, B.~Niu, C.~Lu, \emph{et al.}, ``Mutational landscape and significance across 12 major cancer types," \emph{Nature}, vol. 502, pp. 333-339, 2013.

\bibitem{Szczurek_13}
E.~Szczurek, N.~Misra, and M.~Vingron, ``Synthetic sickness or lethality points at candidate combination therapy targets in glioblastoma," \emph{Int J Cancer}, vol. 133, pp. 2123-2132, 2013.

\bibitem{Wang_X}
X.~Wang and R.~Simon, ``Identification of potential synthetic lethal genes to \emph{p53} using a computational biology approach," \emph{BMC Med Genomics}, vol. 6, 30, 2013.

\bibitem{Jerby-Arnon}
L.~Jerby-Arnon, N.~Pfetzer, Y.~Y. Waldman,  L.~McGarry, D.~James, E.~Shanks, \emph{et al.}, ``Predicting cancer-specific vulnerability via data-driven detection of synthetic lethality," \emph{Cell}, vol. 158, pp. 1199-1209, 2014.

\bibitem{Vaske}
C.~J. Vaske, S.~C. Benz, J.~Z. Sanborn, D.~Earl, C.~Szeto, J.~Zhu, \emph{et al.}, ``Inference of patient-specific pathway activities from multi-dimensional cancer genomics data using PARADIGM," \emph{Bioinformatics}, vol. 26, pp. i237-i245, 2010.

\bibitem{Akavia}
U.~D. Akavia, O.~Litvin, J.~Kim, F.~Sanchez-Garcia, D.~Kotliar, H.~C. Causton, \emph{et al.}, ``An integrated approach to uncover drivers of cancer," \emph{Cell}, vol. 143, pp. 1005-1017, 2010.

\bibitem{Bashashati}
A.~Bashashati, G.~Haffari, J.~Ding, G.~Ha, K.~Lui, J.~Rosner, \emph{et al.}, ``DriverNet: uncovering the impact of somatic driver mutations on transcriptional networks in cancer," \emph{Genome Biol}, vol. 13, R124, 2012.

\bibitem{Zhang_12}
S.~Zhang, C.~C. Liu, W.~Li, H.~Shen, P.~W. Laird, and X.~J. Zhou, ``Discovery of multi-dimensional modules by integrative analysis of cancer genomic data," \emph{Nucleic Acids Res}, vol. 40, pp. 9379-9391, 2012.

\bibitem{Li_12}
W.~Li, S.~Zhang, C.~C. Liu, and X.~J. Zhou, ``Identifying multi-layer gene regulatory modules from multi-dimensional genomic data," \emph{Bioinformatics}, vol. 28, pp. 2458-2466, 2012.

\bibitem{Mo_13}
Q.~Mo, S.~Wang, V.~E. Seshan, A.~B. Olshen, N.~Schultz, C.~Sander, \emph{et al.}, ``Pattern discovery and cancer gene identification in integrated cancer genomic data," \emph{Proc Natl Acad Sci}, vol. 110, pp. 4245-4250, 2013.

\bibitem{Masica}
D.~L. Masica and R.~Karchin, ``Correlation of somatic mutation and expression identifies genes important in human glioblastoma progression and survival," \emph{Cancer Res}, vol. 71, pp. 4550-4561, 2011.

\bibitem{TCGA_13}
The Cancer Genome Atlas Research Network, ``Comprehensive molecular characterization of clear cell renal cell carcinoma," \emph{Nature}, vol. 499, pp. 43-49, 2013.

\bibitem{Cheng}
W.~C. Cheng, I.~F. Chung, C.~Y. Chen, H.~J. Sun, J.~J. Fen, W.~C. Tang, \emph{et al.}, ``DriverDB: an exome sequencing database for cancer driver gene identification," \emph{Nucleic Acids Res}, vol. 42, pp. D1048-D1054, 2014.

\bibitem{Kristensen}
V.~N. Kristensen, O.~C. Lingj${\ae}$rde, H.~G. Russnes, H.~K. Vollan, A.~Frigessi, and A.~L. B${\o}$rresen-Dale, ``Principles and methods of integrative genomic analyses in cancer," \emph{Nat Rev Cancer}, vol. 14, pp. 299-313, 2014.

\bibitem{Liu2015}
Z.~Liu and S.~Zhang, ``Tumor characterization and stratification by integrated molecular profiles reveals essential pan-cancer features," \emph{BMC Genomics}, vol. 16, 503, 2015.

\bibitem{Ciriello_13}
G.~Ciriello, M.~L. Miller, B.~A. Aksoy, Y.~Senbabaoglu, N.~Schultz, C.~Sander, ``Emerging landscape of oncogenic signatures across human cancers," \emph{Nat Genet}, vol. 45, pp. 1127-1133, 2013.

\bibitem{Lawrence_14}
M.~S. Lawrence, P.~Stojanov, C.~H. Mermel, J.~T. Robinson, L.~A. Garraway, T.~R. Golub, \emph{et al.}, ``Discovery and saturation analysis of cancer genes across 21 tumor types," \emph{Nature}, vol. 505, pp. 495-501, 2014.

\end{thebibliography}
\end{document}